\documentclass[aps,prb,amsmath,amssymb,twocolumn,superscriptaddress,showpacs,floatfix]{revtex4-1}

\usepackage{graphicx}
\usepackage{dcolumn}
\usepackage{bm}

\usepackage{amsmath}
\usepackage{amssymb}

\DeclareMathOperator{\Ec}{\mathit E_{\mathrm c}}
\DeclareMathOperator{\Ji}{\mathit J_{\mathrm{sd}}}
\DeclareMathOperator{\rv}{\mathbf r}

\DeclareMathOperator{\Ef}{E_\mathrm F}
\DeclareMathOperator{\epseff}{\varepsilon_\mathrm{eff}}

\begin{document}

\title{Competition of the Coulomb and hopping based exchange interactions in granular magnets}

\author{O.~G.~Udalov}
\affiliation{Department of Physics and Astronomy, California State University Northridge, Northridge, CA 91330, USA}
\affiliation{Institute for Physics of Microstructures, Russian Academy of Science, Nizhny Novgorod, 603950, Russia}
\author{I.~S.~Beloborodov}
\affiliation{Department of Physics and Astronomy, California State University Northridge, Northridge, CA 91330, USA}

\date{\today}

\pacs{75.50.Tt 75.75.Lf	75.30.Et 75.75.-c}

\begin{abstract}
We study exchange coupling due to the interelectron Coulomb interaction
between two ferromagnetic grains embedded into insulating matrix.
This contribution to the exchange interaction complements the contribution
due to virtual electron hopping between the grains. We show that the Coulomb and
the hopping based exchange interactions are comparable. However, for most system parameters
these contributions have opposite signs and compete with each other.
In contrast to the hopping based exchange interaction the Coulomb based exchange
is inversely proportional to the dielectric constant of the insulating matrix $\varepsilon$.
The total intergrain exchange interaction has a complicated dependence on the
dielectric permittivity of the insulating matrix. Increasing $\varepsilon$ one can observe
the ferromagnet-antiferromagnet (FM-AFM) and AFM-FM transitions.
For certain parameters no transition is possible, however even in this case the exchange interaction
has large variations, changing its value by three times with increasing the matrix dielectric constant.
\end{abstract}

\maketitle
\section{Introduction}

Granular metals posses complicated physics involving size and charge quantization effects which
interplay with complicated morphology of these systems~[\onlinecite{Halperin1986,Bel2007review,Glazman2002, Aubin2011,Jaeger2001,Pereira2013,Varlamov2005,Tschersich2003,Efetov1980,Kravets2009}]. Many-body effects play crucial role in granular metals. Electronic and thermal transport properties of granular metals are broadly studied both theoretically and experimentally. These properties are defined by conduction electrons in the systems~[\onlinecite{Bel2007review}]. The
situation becomes more complicated in granular magnets with magnetic metallic grains being
embedded into insulating matrix~[\onlinecite{Bel2007,Fujimori1998,Maekawa1998,Chien1992}]. The
magnetic state of granular magnets is defined by three main interactions: magnetic anisotropy
of a single grain, magneto-dipole interaction between ferromagnetic (FM) grains
and the intergrain exchange interaction. Magnetic properties of granular magnets
were studied in many papers. Numerous papers were devoted to magnetic anisotropy and magneto-dipole interaction~[\onlinecite{Chien1988,Chien1991,Ayton95,Ravichandran96,Djurberg97,Sahoo03,Kechrakos98,Grady1998}].
Much less is known about the exchange interaction between
magnetic grains~[\onlinecite{Freitas2001,Sobolev2012,Hembree1996,Lutz1998,Bel2016ExGr}].
The influence of the intergrain exchange coupling on the magnetic state of the whole
granular magnet are currently understood, however the microscopic picture of the intergrain exchange
interaction is still missing. Note that the intergrain exchange coupling is related to the
conduction electrons. The theory of such a coupling extends the theory of conduction electrons in granular metals.

In most experimental studies the exchange coupling between magnetic grains
was explained using Slonczewski model~[\onlinecite{Slonczewski1989}],
developed for magnetic tunnel junctions (MTJ). Usually, the coupling between grains
was estimated using this model by taking into account the grains surface area.
Recently, it was shown that the intergrain coupling differs from the exchange
coupling in MTJ~[\onlinecite{Bel2016ExGr}]. In granular system the exchange coupling depends
not only on the distance between the grains and on the insulating matrix barrier, but also
on the dielectric properties of the matrix. Such an effect appears due to charge quantization
and the Coulomb blockade effects in FM nanograins.

The intergrain exchange coupling studied in the past was due to virtual electron
hopping between the grains and can be associated with the kinetic energy in the system Hamiltonian.
However, it is known that the many-body Coulomb interaction
also leads to the magnetic exchange interaction~[\onlinecite{Landau3,Vons}].
Recently, the Coulomb based exchange interaction was considered in MTJ~[\onlinecite{Beloborodov2016ExLayer}].
It was shown that this contribution to the magnetic interaction between magnetic leads
separated by the insulating layer is comparable and even larger than the hopping based exchange coupling.

In this paper we consider a competition of the Coulomb and the hopping based exchange coupling
in the system of two spherical magnetic grains embedded into insulating matrix. In contrast
to the layered system the screening of the Coulomb interaction in granular system is different
due to finite grain sizes. This leads to the appearance of additional terms in the total
exchange interaction between grains. Also, the hopping based exchange interaction
in granular and layered systems is different. Thus, the competition of hopping and Coulomb
based exchange interaction in granular system results in essentially different total coupling.

In Ref.~[\onlinecite{Beloborodov2016ExLayer}] it was shown that the Coulomb based coupling strongly
depends on the insulator dielectric constant. For granular system
both the hopping and the Coulomb based exchange depends on the matrix dielectric susceptibility.

In this paper we calculate the Coulomb based exchange interaction
between FM nanograins and study the competition between two mechanisms of exchange interaction.

The paper is organized as follows. In Sec.~\ref{Sec:model} we introduce
the model for granular system. In Secs.~\ref{Sec:Wf} and \ref{Sec:ExHop} we underline the main
results for the hopping based exchange coupling in granular systems. In Sec.~\ref{Sec:ExCoul} we
calculate the inter-electron Coulomb interaction and the intergrain exchange coupling.
We discuss and compare the Coulomb and the hopping based exchange interaction in Sec.~\ref{Sec:Discuss}.
Finally, we discuss validity of our theory in Sec.~\ref{Sec:Val}.
\begin{figure}
\includegraphics[width=1\columnwidth]{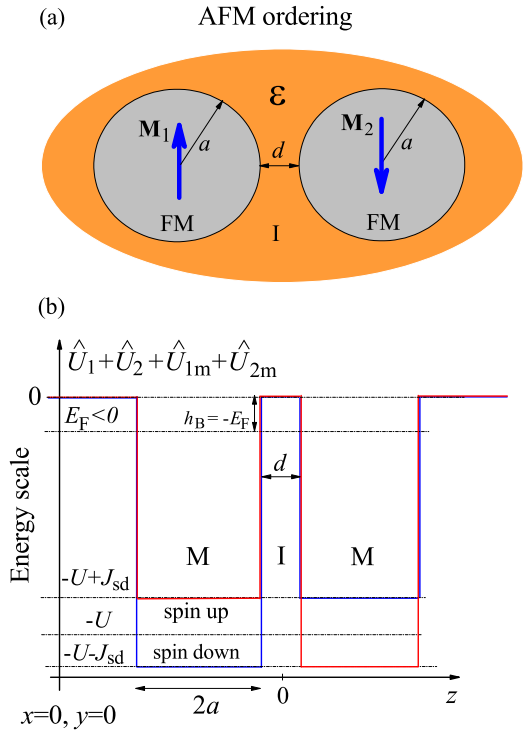}
\caption{(Color online) (a) Two FM metallic grains with radius $a$ and intergrain distance $d$ embedded
into insulating matrix with dielectric constant $\varepsilon$. $\mathbf M_{1,2}$ stands for grain
magnetic moment. (b) Schematic picture of potential energy profiles
for electron with spin ``up'' (red line) and ``down'' (blue line)
states for AFM configuration of leads magnetic moments $\mathbf M_{1,2}$.
Red and blue lines are slightly shifted with respect to each other for better presentation.
Zero energy corresponds to the top of energy barrier for electrons in the insulator.
Symbols FM and I stand for FM metal and insulator, respectively.
All other notations are defined in the text. } \label{Fig:EnExp}
\end{figure}

\section{The model}\label{Sec:model}

We consider two identical FM grains with radius $a$ (see Fig.~\ref{Fig:EnExp}).
The Hamiltonian describing delocalized electrons in the system can be written as follows
\begin{equation}\label{Eq:HamIn}
\begin{split}
&\hat H=\hat H_0+\hat H_\mathrm C,
\end{split}
\end{equation}
where the single particle Hamiltonian $\hat H_0=\sum_i (\hat W_\mathrm k(\mathbf r_i)+\hat U_1(\mathbf r_i)+\hat U_2(\mathbf r_i)+\hat H_{1\mathrm{m}}(\mathbf r_i)+\hat H_{2\mathrm{m}}(\mathbf r_i))$ has the kinetic energy $\hat W_\mathrm k$, the
potential profiles of grains $\hat U_{1,2}$ and the exchange interaction between conduction electrons and ions $\hat H_{1,2\mathrm{m}}$~[\onlinecite{Vons}] in each grain. $\hat H_{\mathrm{C}}$ is the Coulomb interaction between electrons.

We assume that the single particle potential energy is $\hat U_{i}=-U\Pi_i$, where $\Pi_{i}=1$ inside grain ($i$) and $\Pi_{i}=0$ outside grain ($i$). We consider only
FM and AFM collinear configurations of the grains magnetizations $\mathbf{M}_{1,2}$. According to Vonsovskii s-d model the ions influence the delocalized electrons through creation of spin-dependent single particle potential of magnitude $\hat H_{1,2\mathrm{m}}^\mathrm{sp}(\rv_i)=-J_\mathrm{sd}\hat \sigma_z M_{1,2}\Pi_{1,2}$; where
$M_{1,2}$ takes only two possible values $\pm 1$.

Note that we choose the zero energy level at the top of the insulating barrier
(see Fig.~\ref{Fig:EnExp}). This leads to the negative Fermi level, $E_\mathrm F<0$.

We introduce a single particle Hamiltonian for each separate grain,
$\hat H^{\mathrm{g}}_{1,2}=\hat W_\mathrm k+\hat U_{1,2}+\hat H_{1,2\mathrm{m}}$,
with the eigenfunctions $\psi^s_i$ in the grain (1) and $\phi^s_j$ in the grain (2). The subscript $i$ stands for orbital state and the superscript $s$ denotes the spin state in a local spin coordinate system related to magnetization of corresponding grain.
Due to grains symmetry the wave functions are symmetric $\psi^s_i(x,y,z)=\phi^s_i(x,y,-z)$. The
energies of these states are $\epsilon_{1i}^s=\epsilon_{2i}^s=\epsilon_i^s$.

The creation and annihilation operators in grain (1) are
$\hat a^{s+}_i$ and $\hat a^{s}_i$, and in grain (2) are
$\hat b^{s+}_i$ and $\hat b^{s}_i$. The total number of electrons is given
by the operators $\hat n$ and $\hat m$ in grain (1) and (2), respectively.
The whole system is neutral. The total charge of ions in each grain is $-en_0$.

We introduce the zero-order many-particle wave
functions $\Psi^\mathrm{AFM}_0$ and $\Psi^\mathrm{FM}_0$ for
AFM and FM configurations of leads magnetic moments $\mathbf{M}_{1,2}$.
These wave functions describe the non-interacting FM grains ($d\to\infty$).
All states $\psi^s_i$ and $\phi^s_j$ with
energies $\epsilon^s_i<E_\mathrm F$ are filled and all states above $E_\mathrm F$ are empty (we consider the limit of zero temperature). The wave functions of coupled grains, when $d$ is finite,
are denoted as $\Psi^\mathrm{FM}$ and $\Psi^\mathrm{AFM}$ for FM and AFM configurations,
respectively.

We split the Coulomb interaction operator into two parts,
$\hat H_\mathrm C=\hat H_\mathrm{dC}+\hat H_\mathrm{iC}$.
Here $\hat H_\mathrm{dC}$ describes direct Coulomb interaction of electrons in the grains.
It has the form~[\onlinecite{Bel2007review,Glazman2002}]
\begin{equation}\label{Eq:DirCoul}
\hat H_\mathrm{dC}=\Ec (\hat n-n_0)^2+\Ec (\hat m-n_0)^2+\frac{e^2}{C_\mathrm m}(\hat n-n_0)(\hat m-n_0),
\end{equation}
where $E_\mathrm c=e^2/(8\pi\varepsilon_0\epseff a)$ is the grain charging energy in SI units with $\epseff$ being the effective dielectric constant of the surrounding media. In general $\epseff$ can differ from the dielectric constant $\varepsilon$ of the insulating matrix.
In granular magnets the effective dielectric constant depends on properties of insulating matrix and grains~[\onlinecite{Bel2007review}]. In inhomogeneous systems, such as
layers of grains located on top of insulating  substrate, the charging energy,
$E_\mathrm c$, is a complicated function depending on the grain density, dielectric properties of
the substrate and geometrical factors~[\onlinecite{Bel2014ME1,Bel2014GFE2}]. In Eq.~(\ref{Eq:DirCoul}),
$C_\mathrm m$ is the mutual grains capacitance.

The second part of the Coulomb interaction describes the indirect spin-dependent
Coulomb interaction - the
exchange interaction~[\onlinecite{Landau3}]
\begin{equation}\label{Eq:ExHam}
\hat H^\mathrm{}_\mathrm{iC}=-\sum_{i,j,s} U^{\mathrm{s}}_{ij}\hat a^{s+}_i\hat a^{s}_i\hat b^{s'+}_j\hat b^{s'}_j,
\end{equation}
with
\begin{equation}\label{Eq:ExMatEl}
\begin{split}
&U^{s}_{ij}=\!\int\!\!\int d^3\mathbf r_1 d^3 \mathbf r_2 \psi_i^{s*}(\rv_1)\phi^{s'}_j(\rv_1)\hat U_\mathrm C \psi^s_i(\rv_2)\phi^{s'*}_j(\rv_2).\\
\end{split}
\end{equation}
Here $s'=s$ for FM and $s'=-s$ for AFM configuration of grain magnetic moments;
and $\hat U_\mathrm C$ is the operator of the Coulomb interaction between two electrons.
In Eq.~(\ref{Eq:ExHam}) we keep only diagonal elements of the indirect Coulomb interaction with
repeating indexes. We do this assuming that electron wave functions have random phases due to scattering on impurities. In this case only matrix elements with repeating indices survive. Also we omit the indirect Coulomb interaction
between conduction electrons in the same grain. On one hand this contribution does not
produce any interaction between grains and on the other hand it leads to spin subband
splitting which is much smaller than the s-d interaction (and may be incorporated into constant $\Ji$).

Recently the exchange interaction between magnetic grains was considered using
the Hamiltonian $\hat H_0+\hat H_\mathrm{dC}$~[\onlinecite{Bel2016ExGr}]. However, later
it was shown~[\onlinecite{Beloborodov2016ExLayer}] that the indirect Coulomb interaction may also lead to magnetic coupling between the FM contacts.
In particular, this was demonstrated for MTJ with infinite leads.
The indirect Coulomb based interlayer exchange interaction was found to be comparable with hopping
based exchange interaction. In the present paper we calculate the intergrain exchange
interaction based on the indirect Coulomb interaction of electrons,
$\hat H^\mathrm{}_\mathrm{iC}$. We denote the hopping based exchange interaction
as $H^\mathrm{ex}_\mathrm h$. It is given by the following expression
\begin{equation}\label{Eq:ExHopGen}
H^\mathrm{ex}_\mathrm h=\langle\Psi^\mathrm{AFM}|\hat H_0+\hat H_\mathrm{dC}|\Psi^\mathrm{AFM}\rangle-\langle\Psi^\mathrm{FM}|\hat H_0+\hat H_\mathrm{dC}|\Psi^\mathrm{FM}\rangle.
\end{equation}
The contribution to the exchange coupling from the indirect Coulomb interaction is given by
\begin{equation}\label{Eq:ExCoulGen}
H^\mathrm{ex}_\mathrm{iC}=\langle\Psi^\mathrm{AFM}_0|\hat H^\mathrm{}_\mathrm{iC}|\Psi^\mathrm{AFM}_0\rangle-\langle\Psi^\mathrm{FM}_0|\hat H^\mathrm{}_\mathrm{iC}|\Psi^\mathrm{FM}_0\rangle.
\end{equation}
For Coulomb based exchange interaction it is enough to average the
operator over the ground state. The total exchange interaction is defined as follows
\begin{equation}\label{Eq:ExTot}
H^\mathrm{ex}=H^\mathrm{ex}_\mathrm h+H^\mathrm{ex}_\mathrm{iC}.
\end{equation}

\section{Single grain wave functions}\label{Sec:Wf}

Consider single spherical metallic grain with radius $a$. We will follow the approach and notations of Ref.~[\onlinecite{Bel2016ExGr}]. In the absence of spin-orbit interaction the spin and the spatial parts
of wave functions are separated. The spin parts are $(1~0)^T$ and $(0~1)^T$
for the spin up and spin down states, respectively. We introduce the following coordinates: $z$ is along the line connecting grain centres;
$z=0$ is the symmetry point between the grains; $x$ and $y$ are perpendicular to $z$, $r_\perp=\sqrt{x^2+y^2}$. Grains surfaces are close to each other around point $(x,y,z)=0$. In general, the wave functions are the spherical waves with quantum numbers ($m,n,l$). For $d \ll a$ and $S_\mathrm c=\pi a/\varkappa_0 \ll \pi a^2$ ($\varkappa_0=\sqrt{-2m_\mathrm eE_\mathrm F/\hbar^2}$ is the inverse characteristic length scale of electron wave function decay inside the insulator) we approximate the electron wave functions in the vicinity of grain surfaces with plane waves. We change quantum numbers ($m,n,l$) with $(k_x,k_y,k_z)$. In the vicinity of grains contact area we use the following expressions for wave functions
\begin{equation}\label{Eq:WaveFuncR5}
\begin{split}
\psi_\mathbf k^s(z,r_\perp)\approx\frac{\tau_\mathbf k^s}{\sqrt{\Omega}}\exp\left(-\varkappa_\mathbf k^s\left(\frac{d}{2}+ z+\frac{r_\perp^2}{2a}\right)\right)e^{i\mathbf k_\perp \mathbf r_\perp},\\
\phi_\mathbf k^s(z,r_\perp)\approx\frac{\tau^s_\mathbf k}{\sqrt{\Omega}}\exp\left(-\varkappa_\mathbf k^s\left(\frac{d}{2}
- z+\frac{r_\perp^2}{2a}\right)\right)e^{i\mathbf k_\perp \mathbf r_\perp}.
\end{split}
\end{equation}
This expression is valid in the insulator region outside the grains. Here $\tau_\mathbf k^s =\frac{2k_z}{k_z+i\varkappa^s_\mathbf k}$ is the amplitude of the transmitted electron wave,
$\mathbf k_{\perp}=(k_x,k_y,0)$, $\mathbf r_{\perp}=(x,y,0)$, $\Omega=4\pi a^3/3$ and
$\varkappa^s_\mathbf k=\sqrt{2m_\mathrm e(U-s\Ji-\hbar^2k^2_z/(2m_\mathrm e))/\hbar^2}$
is the inverse decay length written in new notations. We neglect
the surface interference effect and the penetration of electron wave function beyond the
grain in determining the normalization factor.

Below we will use the symbols $i$ and $j$ (instead of $\mathbf k$) to describe a set of quantum numbers characterizing the
orbital motion of electrons. The overlap of wave functions of electrons $i$ and $j$ located in different grains
exists only between the grains in a small region in the vicinity of $r_\perp=0$.
The in-plane area (($x,y$)-plane) of the overlap region is $S^{ij}_\mathrm c=\pi (\lambda^{ij}_\perp)^2$, where $\lambda^{ij}_\perp=\sqrt{2a/(\varkappa_i+\varkappa_j)}$. The introduced above area,
$S_\mathrm c=\pi\lambda^2_\perp$, is the contact area for electrons at the Fermi level (size $\lambda_\perp=\sqrt{a/\varkappa_0}$).

For electron wave functions inside the grains we obtain
\begin{equation}\label{Eq:WaveFuncR6}
\begin{split}
&\psi^s_\mathbf k(z,r_\perp)\approx\frac{e^{ik_{z}\left(\frac{d}{2}+ z+\frac{r_\perp^2}{2a}\right)}+\xi^s_\mathbf k e^{-ik_z\left(\frac{d}{2}+ z+\frac{r_\perp^2}{2a}\right)}}{\sqrt{\Omega}}e^{i\mathbf k_\perp \mathbf r_\perp},\\
&\phi^s_\mathbf k(z,r_\perp)\approx\frac{e^{ik_{z}\left(\frac{d}{2}- z+\frac{r_\perp^2}{2a}\right)}+\xi^s_\mathbf k e^{-ik_z\left(\frac{d}{2}- z+\frac{r_\perp^2}{2a}\right)}}{\sqrt{\Omega}}e^{i\mathbf k_\perp \mathbf r_\perp},
\end{split}
\end{equation}
with $\xi^s_{\mathbf k} = \frac{k_z-i\varkappa^s_{\mathbf k }}{k_z+i\varkappa^s_{\mathbf k }}$. Below we will use Eqs.~(\ref{Eq:WaveFuncR5}) and (\ref{Eq:WaveFuncR6}) to calculate exchange interaction between the grains.

\section{Hopping based exchange interaction}\label{Sec:ExHop}

This mechanism was considered for grains in Ref.~[\onlinecite{Bel2016ExGr}]. We split the expression for the hopping based exchange interaction into two parts
\begin{equation}\label{Eq:ExFin}
H^\mathrm{ex}_\mathrm h=H^\mathrm{ex}_\mathrm{h0}-H^\mathrm{ex}_\mathrm{h\varepsilon},
\end{equation}
where
\begin{equation}\label{Eq:ExHop0}
\begin{split}
&H^{\mathrm{ex}}_{\mathrm{h0}}\!=\frac{\pi a}{(2\pi)^2\varkappa_0}\!\sum_{s}\!\int^{k_\mathrm F^s}_0\!\!dk (\!(k^s_\mathrm F)^2-k^2)V^s_{k}-\\
&\!-\frac{ a}{8\pi^2\varkappa_0}\sum_s\left[\int^{k^{-s}_\mathrm F}_{0}\!\!\!\!dk_1\!\!\!\int^{k^s_\mathrm F}_0\!\!\!\!\!\!dk_2 \tilde \delta^s(k_1,k_2)T^{-ss}_{12}(P_{12}^{-ss})^*-\right.\\
&-\left.\int^{k^s_\mathrm{F}}_{0}\!\!\!\!dk_1\int^{k^s_\mathrm{F}}_0\!\!dk_2 \delta^s(k_1,k_2) T^{ss}_{12}(P^s_{12})^*\right].
\end{split}
\end{equation}
and
\begin{equation}\label{Eq:ExHop1}
\begin{split}
&H^{\mathrm{ex}}_{\mathrm{h\varepsilon}}\!=\\&=-\frac{a}{8\pi^2\varkappa_0}\left\{\int^{k^-_\mathrm{max}}_{\sqrt{2\tilde J_\mathrm{sd}}}\!\!\!\!dk_1\int^{k^-_\mathrm{up}}_0\!\!\!\!\!\!dk_2\right.\frac{\tilde\xi^-(k_1,k_2)|T^{--}_{12}|^2}{\frac{\hbar^2(k_1^2-k_2^2-2\tilde J_\mathrm{sd})}{2m_\mathrm e}+\tilde \epsilon_\mathrm c}+\\
&+\!\!\!\int^{k^+_\mathrm{max}}_{0}\!\!\!\!dk_1\!\!\!\int^{k^+_\mathrm{up}}_0\!\!\!\!\!\!dk_2\frac{\tilde\xi^+(k_1,k_2)| T^{++}_{12}|^2}{\frac{\hbar^2}{2m_\mathrm e}(k_1^2-k_2^2+2\tilde J_\mathrm{sd})+\tilde \epsilon_\mathrm c}-\\
&-\!\sum_s\int^{k^s_\mathrm{max}}_{0}\!\!\!\!\int^{\mathrm{min}(k_1,k_\mathrm F^s)}_0\!\!\!\!\!\!dk_1 dk_2\left.\frac{\xi^s(k_1,k_2)|T^{s-s}_{12}|^2}{\frac{\hbar^2(k_1^2-k_2^2)}{2m_\mathrm e}+\tilde \epsilon_\mathrm c}\right\}.
\end{split}
\end{equation}
For simplicity we change all different squares $S_\mathrm c^{ij}$ in the
integrals with characteristic contact area $S_0=\pi a/\varkappa_0$. This change
does not influence the resulting exchange interaction a lot.
We introduce the following functions
\begin{equation}\label{Eq:AuxFunc3}
\tilde \delta^s(k_1,k_2)=\left\{\begin{array}{l} (k^{-s}_\mathrm F)^2-k^2_1,~2s\tilde J_{\mathrm{sd}}+k^2_2<k^2_1,\\
(k^s_\mathrm F)^2-k^2_2,~2s\tilde J_{\mathrm{sd}}+k^2_2>k^2_1,~\end{array}\right.
\end{equation}
\begin{equation}\label{Eq:AuxFunc2}
\delta^s(k_1,k_2)=\left\{\begin{array}{l} (k^s_\mathrm F)^2-k^2_1,~k_2<k_1,\\
(k^s_\mathrm F)^2-k^2_2,~k_1>k_2,~\end{array}\right.
\end{equation}
\begin{equation}\label{Eq:AuxFunc1}
\tilde\xi^s(k_1,k_2)=\left\{\begin{array}{l}(2s\tilde J_\mathrm{sd}+k_1^2-k_2^2),~k_1<k_\mathrm F^s,\\
((k_\mathrm F^{-s})^2-k_2^2),k_1>k_\mathrm F^s,~\end{array}\right.
\end{equation}
\begin{equation}\label{Eq:AuxFunc}
\xi^s(k_1,k_2)=\left\{\begin{array}{l}(k_1^2-k_2^2),~k_1<k_\mathrm F^s,\\
((k_\mathrm F^s)^2-k_2^2),k_1>k_\mathrm F^s,~\end{array}\right.
\end{equation}
and notations
\begin{equation}\label{Eq:AuxFunc4}
k^s_\mathrm{up}=\mathrm{min}(\sqrt{k^2_1+2s\tilde J_\mathrm{sd}} \, ,k_\mathrm F^{-s}).
\end{equation}
\begin{equation}\label{Eq:AuxNot1}
\begin{split}
&k^s_\mathrm{max}=\sqrt{2m_e(U-sJ)/\hbar^2},
\end{split}
\end{equation}
\begin{equation}\label{Eq:AuxNot2}
\begin{split}
&k^s_\mathrm{F}=\sqrt{2m_e(E_\mathrm F+U-sJ)/\hbar^2},
\end{split}
\end{equation}
\begin{equation}\label{Eq:AuxNot3}
\begin{split}
\tilde J_\mathrm{sd}=2m_\mathrm e\Ji/\hbar^2.
\end{split}
\end{equation}
We introduce the charging energy
$\tilde \epsilon_\mathrm c=2E_\mathrm c-e^2/C_\mathrm m$, which can be estimated as
$\tilde \epsilon_\mathrm c=e^2/(8\pi a\epsilon\epsilon_0)$ for $d\approx1$ nm and $a\in\left[1;10\right]$ nm.

The matrix elements $T^s_{12}$, $P^s_{12}$, and $V^s_{k}$ in Eqs.~(\ref{Eq:ExHop0}) and (\ref{Eq:ExHop1}) are given by the following expressions
\begin{equation}\label{Eq:ME7}
\begin{split}
&V^s_{k}=- s\!\Ji\frac{(|\tau^s_{i}|)^2}{(\varkappa_i^s)}e^{-2\varkappa_i^s d},\\
&T^{ss'}_{ij}\!=-(s\!\Ji+U)\frac{\tau^{s*}_{i}\tau^{s'}_{j}(\varkappa_i^{s}\!\!+\varkappa_j^{s'})}{(\!(k_i)^2+\!(\varkappa_j^{s'})^2)}e^{-\varkappa_j^{s'} \!d},\\
&P_{ij}^{ss'}=\frac{\tau^{s*}_{i}\tau^{s'}_{j}(\varkappa_i^{s}\!\!+\varkappa_j^{s'})}{(k_j^2+\!(\varkappa_i^{s})^2)}e^{-\varkappa_i^{s} \!d}+\frac{\tau^{s*}_{i}\tau^{s'}_{j}(\varkappa_i^{s}\!\!+\varkappa_j^{s'})}{(k_i^2+\!(\varkappa_j^{s'})^2)}e^{-\varkappa_j^{s'} \!d}+\\&+\frac{ 2\tau^{s*}_{i}\tau^{s'}_j e^{-(\!\varkappa^{s}_i+\varkappa^{s'}_j\!)\frac{d}{2}}\mathrm{sinh}(\!(\!\varkappa_i^{s}-\varkappa_j^{s'}\!)\frac{d}{2})}{(\varkappa_i^{s}-\varkappa_j^{s'})}.\\
\end{split}
\end{equation}

For semimetal with only one spin subband occupied ($E_\mathrm F<\Ji-U$) we sum in Eqs.~(\ref{Eq:ExHop0}) and (\ref{Eq:ExHop1})
only over the occupied spin subband ($s=$``-'').

\section{Coulomb based exchange interaction} \label{Sec:ExCoul}

Integral in Eq.~(\ref{Eq:ExMatEl}) includes the operator of the Coulomb interaction $\hat U_\mathrm C$. For homogeneous insulator it has the form $\hat U_{\mathrm{C}}=e^2/(4\pi\varepsilon_0\varepsilon |\rv_1-\rv_2|)$, where $\varepsilon$
is the medium effective dielectric constant. In our case the system is
inhomogeneous and the Coulomb interaction is renormalized by screening effects due
to metallic grains.

There are two regions contributing to Eq.~(\ref{Eq:ExMatEl}): 1) The region inside the
FM grains $\Omega_{1}$ ($\Omega_2$) where the Coulomb interaction is effectively
screened and is short-range ~[\onlinecite{Glazman2002,Muzykantskii1997}]
\begin{equation}\label{Eq:CoulLeads}
\begin{split}
\hat U^\mathrm L_\mathrm C=&\frac{\Omega\Delta}{2}\delta(\mathbf r_1-\mathbf r_2)+2E_\mathrm c+\\&+\frac{2E_\mathrm c\lambda_\mathrm{TF}^2}{a}\delta(|\mathrm r_1|-a)+\frac{2E_\mathrm c\lambda_\mathrm{TF}^2}{a}\delta(|\mathrm r_2|-a),
\end{split}
\end{equation}
where $\Delta$ is the mean energy level spacing, $\Omega\Delta=6\pi^2E_\mathrm F/((k_\mathrm F^+)^3+(k_\mathrm F^-)^3)$.
In metals the Coulomb interaction is screened on the length scale of the order of
Thomas-Fermi length, $\lambda_\mathrm{TF}\approx(\sqrt{e^2k_\mathrm F^3/(4\pi\varepsilon_0)E_\mathrm F})^{-1} \approx 0.05$ nm.
The characteristic length scale of the electron density variation is
$\varkappa_0^{-1}\approx0.5$ nm. Thus,
we can use the local approximation for decaying electron
wave functions since $\lambda_\mathrm{TF}\ll\varkappa_0^{-1}$.

The Coulomb based exchange coupling between infinite magnetic
leads was considered in Ref.~[\onlinecite{Beloborodov2016ExLayer}], where it was shown that
the Coulomb interaction inside the FM leads also contributes to the total
interlayer exchange coupling. However, for infinite leads the last three terms
in Eq.~(\ref{Eq:CoulLeads}) disappear. In the present paper we take
into account these terms appearing due to finite grain sizes.

2) The second region contributing to Eq.~(\ref{Eq:ExMatEl}) is the region
between the grains where screening of the Coulomb interaction is
weak and the interaction is long-range.
However, due to metallic grains, the electric field of two interacting
electrons is finite only inside this region. We denote the renormalized Coulomb
interaction inside the insulating layer as $\hat U^\mathrm I_\mathrm C$.

In our model electrons inside the insulator and electrons inside the grains do not interact with each other.

The right hand side of Eq.~(\ref{Eq:ExMatEl}) can be considered as the Coulomb interaction between two effective charges, $\rho_{ij}^{(1)}=e\psi^{s*}_i(\rv)\phi^{s'}_j(\rv)$ and $\rho_{ij}^{(2)}=e\psi^s_i(\rv)\phi^{s'*}_j(\rv)$. Here $s'=s$ for FM
and $s'=-s$ for AFM ordering. One can see that $\rho_{ij}^{(1)}=\rho_{ij}^{(2)*}=\rho_{ij}$.

We can write the matrix elements of the indirect Coulomb interaction as a sum of two terms
\begin{equation}\label{Eq:ExMatEl2}
\begin{split}
&U^{s}_{ij}=L^s_{ij}+I^s_{ij},\\
&L^s_{ij}=\!\int\!\!\int_{\Omega_1+\Omega_2} \!\!d^3\mathbf r_1 d^3 \mathbf r_2 \rho_{ij}(\rv_1)\hat U^\mathrm L_\mathrm C \rho_{ij}^*(\rv_2),\\
&I^s_{ij}=\!\int\!\!\int_{\Omega_\mathrm I} d^3\mathbf r_1 d^3 \mathbf r_2 \rho_{ij}(\rv_1)\hat U^\mathrm{I}_\mathrm C \rho_{ij}^*(\rv_2),
\end{split}
\end{equation}
where $\Omega_{1,2}=\Omega$ is the grain volume
and $\Omega_\mathrm I$ is the volume of the insulating layer.
The index $s$ stands for spin index of electron wave function
in grain (1). The spin state of electron in grain (2) is the same ($s$)
for FM and $-s$ for AFM configuration.
We can split the total Coulomb based exchange interaction into two contributions
\begin{equation}\label{Eq:Ex2}
H^\mathrm{ex}_{\mathrm C}=L^\mathrm{ex}+I^\mathrm{ex}.\\
\end{equation}
Below we consider these two contributions to the Coulomb based exchange interaction separately.

\subsection{Contribution to the exchange interaction due to the insulating region, $I^\mathrm{ex}$}

To calculate the contribution to the exchange interaction due to the insulating
region we will follow the approach of Ref.~[\onlinecite{Beloborodov2016ExLayer}] where exchange coupling
was calculated for MTJ. In this approach the electric field $\mathbf E^{ij}_{1,2}$ created by
effective charges $\rho_{ij}^{(1,2)}$ inside the insulating region was calculated by
taking into account the screening produced by the FM leads. The leads were treated as ideal metal
with zero screening length. The energy of this field (the part corresponding to the mutual interaction) $I_{ij}=(\varepsilon_0\varepsilon)(\int_{\Omega_\mathrm I} d^3r \mathbf E^{ij}_{1}\mathbf E^{ij}_{2})$ gives
the estimate of the matrix element of indirect Coulomb interaction. In MTJ the charges $\rho_{ij}^{(1,2)}$
are periodic functions in the (x,y) plane and decay exponentially along $z$ direction.
In the case of magnetic grains the geometry of the system is more complicated.
We will use the following approximation: the region of interaction of electrons
in states $i$ and $j$ is restricted by the area $S^{ij}_\mathrm c$. The linear size
of this area is much larger than the Fermi length for large enough grains
($\sqrt{\pi a/\varkappa}>1/k_\mathrm F$). In this case we can model the interaction
region as two leads with parallel surfaces neglecting grains curvature.
In the region of interaction we calculate the electric field created
by charges $\rho_{ij}^{(1,2)}$ as if we have the infinite parallel leads.
The matrix element of the interaction is given
by $I_{ij}=(\varepsilon_0\varepsilon/2)(\int_{\tilde\Omega_\mathrm I} d^3r \mathbf E^{ij}_{1}\mathbf E^{ij}_{2})$,
where $\tilde \Omega_\mathrm I$ is the volume restricted by the inequalities $|z|<d/2$, $r_\perp<a\varkappa_{ij}$.
In practice, we multiply the area-normalized matrix elements in
Ref.~[\onlinecite{Beloborodov2016ExLayer}]
by the contact area $S^{ij}_\mathrm c$.
Following Ref.~[\onlinecite{Beloborodov2016ExLayer}] we derive the following expression for the
Coulomb based exchange interaction
\begin{equation}\label{Eq:ExIns1}
I^\mathrm{ex}=\tilde{\tilde I}_\mathrm{ex}-\tilde I_\mathrm{ex}^+-\tilde I_\mathrm{ex}^-,
\end{equation}
where
\begin{equation}\label{Eq:ExIns2}
\begin{split}
 &\tilde{\tilde I}_\mathrm{ex}=-\frac{e^2a}{16\pi^4\varepsilon_0\varepsilon} \int_{0}^{k^+_\mathrm F}\int_{0}^{k^-_\mathrm F}\!\!\! dk_{1}dk_{2} \frac{|(\tau^+_1)^*\tau^-_2|^2}{\varkappa_1^+ +\varkappa_2^-} e^{-d(\varkappa^+_1+\varkappa^-_2)} \times \\ &\times \int_{0}^{k_2^{\mathrm{max}}+k_1^{\mathrm{max}}}q \omega_\mathrm I (q) dq \!\!\int_0^{(k_2^{\mathrm{max}}+k_1^{\mathrm{max}})/2}\!\!\! k\zeta(k,q)dk.
\end{split}
\end{equation}
\begin{equation}\label{Eq:ExIns3}
\begin{split}
&\tilde I^s_\mathrm{ex}= -\frac{e^2}{16\pi^4\varepsilon_0\varepsilon} \int_{0}^{k^s_\mathrm F}\int_{0}^{k_1}\!\!\! dk_{1}dk_{2} \frac{|(\tau^s_1)^*\tau^s_2|^2}{\varkappa_1^s +\varkappa_2^s} e^{-d(\varkappa^s_1+\varkappa^s_2)} \times \\ &\times \int_{0}^{k_2^{\mathrm{max}}+k_1^{\mathrm{max}}}q \omega_\mathrm I (q) dq \!\!\int_0^{(k_2^{\mathrm{max}}+k_1^{\mathrm{max}})/2}\!\!\! k\zeta(k,q)dk.
\end{split}
\end{equation}
The maximum value of perpendicular momenta
are $k_1^{\mathrm{max}}=\sqrt{(k^s_\mathrm F)^2 - k^2_{1z}}$ and
$k_2^{\mathrm{max}}=\sqrt{(k^{s'}_\mathrm F)^2 - k^2_{2z}}$, where $s'=s$
in expression for $k_1^\mathrm{max}$ and $k_2^\mathrm{max}$ in Eq.~(\ref{Eq:ExIns2}),
and $s=$``+'', $s'=$``-'' in Eq.~(\ref{Eq:ExIns3}). We also introduce the following functions
\begin{equation}\label{Eq:InPlaneInt2}
\begin{split}
\zeta (k,q) = \left\{\begin{array}{l}0,~~(\phi_2<\phi_3)~ \mathrm{or}~ (\phi_1<\phi_3),\\
\phi_1-\phi_3,~~\mathrm{otherwise,}\end{array}\right.
\end{split}
\end{equation}
where
\begin{equation}\label{Eq:InPlaneInt3}
\begin{split}
\phi_1(k,q)=\left\{\begin{array}{l}0,~~k>k_1^{\mathrm{max}}+q/2,\\
\frac{\pi+\pi\mathrm{sign}(k_1^{\mathrm{max}}-q/2)}{2},~~k<|k_1^{\mathrm{max}}-q/2|,\\
\mathrm{arccos}\left(\frac{k^2+q^2/4-(k_1^{\mathrm{max}})^2}{qk}\right),~~\mathrm{otherwise}. \end{array}\right.
\end{split}
\end{equation}
\begin{equation}\label{Eq:InPlaneInt4}
\begin{split}
\phi_2(k,q)=\left\{\begin{array}{l}\pi,~~k<k_2^{\mathrm{max}}-q/2,\\
\mathrm{arccos}\left(\frac{k^2+q^2/4-(k_2^{\mathrm{max}})^2}{qk}\right),~~\mathrm{otherwise}. \end{array}\right.
\end{split}
\end{equation}
\begin{equation}\label{Eq:InPlaneInt5}
\begin{split}
\phi_3(k,q)=\pi-\phi_2(k,q).
\end{split}
\end{equation}
The reduced matrix element $\omega_\mathrm I (q)$ is given by the expression
\begin{equation}\label{Eq:IntEnTot}
\omega_\mathrm I (q) =\omega_{\mathrm Ix}(q) + \omega_{\mathrm Iz}(q),
\end{equation}
where
\begin{equation}\label{Eq:IntEn}
\begin{split}
&\omega_{\mathrm Iz}=\left\{(\alpha_1^2+\alpha^2_2)\frac{\mathrm{sinh}(dq)}{q}+\alpha_3^2\frac{\mathrm{sinh}(d\Delta\varkappa)}{q}+2\alpha_1\alpha_2d+\right.\\ &\left.+4\alpha_1\alpha_3\frac{\mathrm{sinh}(\!(\Delta\varkappa+q)d/2)}{\Delta\varkappa+q}+4\alpha_2\alpha_3\frac{\mathrm{sinh}(\!(\Delta\varkappa-q)d/2)}{\Delta\varkappa-q}\!\!\right\},\\
&\omega_\mathrm {Ix}=\left\{(\tilde\alpha_1^2+\tilde\alpha^2_2)\frac{\mathrm{sinh}(dq)}{q}+\tilde\alpha_3^2\frac{\mathrm{sinh}(d\Delta\varkappa)}{q}+2\tilde\alpha_1\tilde\alpha_2d+\right.\\&+4\tilde\alpha_1\tilde\alpha_3\frac{\mathrm{sinh}(\!(\Delta\varkappa+q)d/2)}{\Delta\varkappa+q}\left.+4\tilde\alpha_2\tilde\alpha_3\frac{\mathrm{sinh}(\!(\Delta\varkappa-q)d/2)}{\Delta\varkappa-q}\!\!\right\},\\
\end{split}
\end{equation}
where $\Delta\varkappa=\varkappa^s_1-\varkappa^{s'}_2$ and
functions $\alpha_i$ and $\tilde \alpha_i$ are defined as follows
\begin{equation}\label{Eq:IntEn2}
\begin{split}
&\alpha_1\!=e^{-\frac{qd}{2}}\sigma_2-\frac{e^{(\Delta\varkappa-q)\frac{d}{2}}}{q-\Delta\varkappa},~ \tilde \alpha_1\!=-e^{-\frac{qd}{2}}\sigma_2-\frac{e^{(\Delta\varkappa-q)\frac{d}{2}}}{q-\Delta\varkappa},\\
&\alpha_2\!=e^{-\frac{qd}{2}}\sigma_1+\frac{e^{-(q+\Delta\varkappa)d/2}}{q+\Delta\varkappa},~\tilde \alpha_2\!=e^{-\frac{qd}{2}}\sigma_1-\frac{e^{-(q+\Delta\varkappa)\frac{d}{2}}}{q+\Delta\varkappa},\\
&\alpha_3=\frac{2\Delta\varkappa}{q^2-\Delta\varkappa^2}, ~~\tilde \alpha_3=\frac{-2q}{q^2-\Delta\varkappa^2}.
\end{split}
\end{equation}
The functions $\sigma_{1,2}$ are defined as
\begin{equation}\label{Eq:SurfCharges}
\begin{split}
\sigma_{1(2)}=\frac{\sigma_{1(2)}^0e^{qd}+\sigma_{2(1)}^0}{e^{qd}-e^{-qd}},
\end{split}
\end{equation}
with
\begin{equation}\label{Eq:SurfCharges2}
\begin{split}
\sigma_1^0=\frac{e^{-qd/2}}{q-\Delta \varkappa}\left(e^{(q-\Delta\varkappa)d/2}-e^{-(q-\Delta\varkappa)d/2}\right),\\
\sigma_2^0=\frac{e^{-qd/2}}{q+\Delta \varkappa}\left(e^{-(q+\Delta\varkappa)d/2}-e^{(q+\Delta\varkappa)d/2}\right).
\end{split}
\end{equation}

\subsection{Contribution to the exchange interaction due to grains, $L^\mathrm{ex}$}

In this region the operator of Coulomb interaction
is defined in Eq.~(\ref{Eq:CoulLeads}). The operator consists of four terms.
The last three terms contribute only in the case of nanoscale grains. These terms vanish
for infinite metallic leads.

First, we consider the last two terms describing
single particle potential uniformly distributed over the grain surface.
This potential is zero inside the grain. Consider the interaction between
an electron in some state $\psi_i^s$ located in the left grain and an electron in state $\phi^s_j$ located in the right grain.
 Consider the interior of the right grain.
The charge $\rho_{ij}$ is non-zero only in the small area $S^{ij}_\mathrm c$ in the (x,y) plane
and penetrates into the grain by the distance $\varkappa^{-1}$. Therefore the potential $\frac{2E_\mathrm c\lambda_\mathrm{TF}^2}{a}\delta(|\mathrm r_2|-a)$ interacts with the charge $\rho_{ij}$ only
in the small area of the surface $S^{ij}_\mathrm c \ll 4\pi a^2$.
Therefore this potential gives a small contribution to the intergrain
exchange interaction in comparison to the contribution coming from the first term of Eq.~(\ref{Eq:CoulLeads}), $\frac{\Omega\Delta}{2}\delta(\mathbf r_1-\mathbf r_2)$. The direct calculations show that the small parameter is
$(a\varkappa_0)^{-1}(ak_\mathrm F)^{-1}(E_\mathrm F/E_\mathrm c) \ll 1$. For this reason we neglect the last two
terms in Eq.~(\ref{Eq:CoulLeads}).

The matrix element calculated using the second term in Eq.~(\ref{Eq:CoulLeads}) is given by
\begin{equation}\label{Eq:MatElCoulEc}
\begin{split}
&2E_\mathrm c\int\int_{\Omega_1+\Omega_2} d^3r_1 d^3r_2 \rho_{ij}(\mathbf r_1)\rho_{ij}^*(\mathbf r_2)=2E_\mathrm c|\tau^s_i|^2|\tau^{s'}_j|^2\times\\
&\times\frac{(\varkappa^s_i+\varkappa^{s'}_j)^2}{\Omega^2}\left|\frac{e^{-\varkappa^{s'}_jd}S^{s'}_j\mathrm{Sinc}(q_x\lambda_\perp^j)\mathrm{Sinc}(q_y\lambda_\perp^j)}{(k^s_i)^2+(\varkappa_j^{s'})^2}\right.+\\
&\left.\frac{e^{-\varkappa^{s}_id}S^s_i\mathrm{Sinc}(q_x\lambda_\perp^i)\mathrm{Sinc}(q_y\lambda_\perp^i)}{(k^{s'}_j)^2+(\varkappa_i^{s})^2}\right|^2.
\end{split}
\end{equation}
Here $S^s_i=\pi a/\varkappa^s_i$ is the surface area and
$\lambda_\perp^i=\sqrt{S^s_i}$ is the linear size, and
$\mathbf q=\mathbf k_{1\perp}-\mathbf k_{2\perp}$ is the momentum.
The contribution to the intergrain exchange coupling due to this matrix element is
\begin{equation}\label{Eq:ExchContrCoulEc}
\begin{split}
&L^\mathrm{ex}_{E_\mathrm c}=\frac{-e^2}{64\pi^3\varepsilon\varepsilon_0}\sum_s\!\int_0^{k^s_\mathrm F}\!\!\int_0^{k^s_\mathrm F}\! dk_1dk_2 |\tau^s_1|^2|\tau^{s}_2|^2\delta(k_1,k_2)\times\\
&\times(\varkappa^s_1+\varkappa^{s}_2)^2\left\{\frac{e^{-2\varkappa^{s}_1d}}{(k_2^2+(\varkappa_1^{s})^2)^2\varkappa_1^{s}}+\frac{e^{-2\varkappa^{s}_2d}}{(k_1^2+(\varkappa_2^{s})^2)^2\varkappa_2^{s}}+\right.\\
&\left.\frac{e^{-(\varkappa^{s}_1+\varkappa^{s}_2)d}}{(k_2^2+(\varkappa_1^{s})^2)(k_1^2+(\varkappa_2^{s})^2)\mathrm{max}(\varkappa_1^{s},\varkappa_2^{s})}\right\}-\\
&-\frac{e^2}{32\pi^3\varepsilon\varepsilon_0}\!\int_0^{k^+_\mathrm F}\!\!\int_0^{k^-_\mathrm F}\! dk_1dk_2 |\tau^+_1|^2|\tau^{-}_2|^2\tilde\delta(k_1,k_2)\times\\
&\times(\varkappa^+_1+\varkappa^{-}_2)^2\left\{\frac{e^{-2\varkappa^{+}_1d}}{(k_2^2+(\varkappa_1^{+})^2)^2\varkappa_1^{+}}+\frac{e^{-2\varkappa^{-}_2d}}{k_1^2+(\varkappa_2^{-})^2)^2\varkappa_2^{-}}+\right.\\
&\left.\frac{e^{-(\varkappa^{+}_1+\varkappa^{-}_2)d}}{(k_2^2+(\varkappa_1^{+})^2)(k_1^2+(\varkappa_2^{-})^2)\mathrm{max}(\varkappa_1^{+},\varkappa_2^{-})}\right\}.
\end{split}
\end{equation}
The first term in Eq.~(\ref{Eq:CoulLeads}) gives the following contribution to the intergrain exchange interaction
\begin{equation}\label{Eq:ExchContrCoulDelta}
\begin{split}
&L^\mathrm{ex}_{\mathrm{loc}}=\frac{-3a(U+E_\mathrm F)}{2^6\pi((k_\mathrm F^+)^3+(k_\mathrm F^-)^3))}\sum_{s,s'}\gamma(s,s')\times\\&\times\int_0^{k_\mathrm F^s}\!\!\int_0^{k_\mathrm{F}^{s}}\!\!dk_1 dk_2((k_\mathrm F^{s'})^2-k_2^2)((k_\mathrm F^s)^2-k_1^2)\times\\&\times\left\{\frac{e^{-2d\varkappa^s_{1}}|\tau^s_{1}|^2}{\varkappa^s_{1}}\left(\frac{1+|r^{s'}_{2}|^2}{2\varkappa_1^s}+\mathrm{Re}\left(\frac{(r^{s'}_{2})^*}{\varkappa^s_{1}+ik_2}\right)\right)+\right.\\
&\left. \frac{|\tau^{s'}_{2}|^2e^{-2d\varkappa^{s'}_{2}}}{\varkappa^{s'}_{2}}\left(\frac{1+|r^{s}_{1}|^2}{2\varkappa_2^{s'}}+\mathrm{Re}\left(\frac{(r^{s}_1)^*}{\varkappa^{s'}_2+ik_1}\right)\right)\right\},\\
\end{split}
\end{equation}
we introduce the function
\begin{equation}
\gamma(s,s')=\left\{\begin{array}{l}1,~s=s',\\ -1,~ s\ne s'.\end{array}\right.
\end{equation}

\subsection{Total exchange interaction}

The total intergrain exchange interaction is given by the following expression
\begin{equation}\label{Eq:TotExFin}
H^\mathrm{ex}=H^\mathrm{ex}_\mathrm{h0}+L^\mathrm{ex}_{\mathrm{loc}}+H^\mathrm{ex}_\mathrm{h\varepsilon}+ I^\mathrm{ex}+L^\mathrm{ex}_{E_\mathrm c},
\end{equation}
where term $H^\mathrm{ex}_\mathrm{h0}$ is given by Eq.~(\ref{Eq:ExHop0}), $L^\mathrm{ex}_{\mathrm{loc}}$ by Eq.~(\ref{Eq:ExchContrCoulDelta}), $H^\mathrm{ex}_\mathrm{h\varepsilon}$ by Eq.~(\ref{Eq:ExHopEps}), $I^\mathrm{ex}$ by Eqs.~(\ref{Eq:ExIns1}-\ref{Eq:ExIns3}) and $L^\mathrm{ex}_{E_\mathrm c}$ by Eq.~(\ref{Eq:ExchContrCoulEc}).

\section{Discussion of results}\label{Sec:Discuss}

There are several contributions to the intergrain exchange interaction
in Eq.~(\ref{Eq:TotExFin}). These contributions have different physical nature
and different dependencies on system parameters. In this section we
will discuss these contributions and compare
the intergrain exchange coupling with the interlayer exchange
coupling in MTJ.

\subsection{Granular magnets}

First, we discuss the influence of intergrain exchange interaction
on properties of granular magnets with many grains forming an
ensemble of interacting nanomagnets. The exchange interaction between the
grains leads to the formation of long-range magnetic order appearing below
a certain temperature~[\onlinecite{Glazman2002, Freitas2001, Sobolev2012, Hembree1996}],
which is called the ordering temperature $T_\mathrm{ord}$.
For Ising model~[\onlinecite{Sobolev2012, Munakata2009}] the ordering temperature in
granular magnets with FM coupling is related to the intergrain exchange interaction as $T_\mathrm{ord}=z_\mathrm n H^\mathrm{ex}$, where $z_\mathrm n =6$ is the coordination number for three dimensional cubic lattice.
Below we will plot the exchange interaction multiplied by the
coordination number, $z_\mathrm n=6$, to show the
temperature where coupling overcomes temperature fluctuations.

Note that we do not consider the intergrain
magneto-dipole (MD) interaction~[\onlinecite{Kechrakos98, Grady1998, Ayton95, Ravichandran96, Mamiya99, Djurberg97, Sahoo03}], which competes with the exchange interaction and leads to the
formation of super spin glass state. The influence of MD interaction on the magnetic
state of GFM was discussed in Refs.~[\onlinecite{Ayton95, Ravichandran96, Mamiya99, Djurberg97, Sahoo03}].

\subsection{Comparison with layered systems}

Both, the hopping and the Coulomb based exchange coupling were considered
for layered structures such as MTJ in the past.
There are at least three essential differences between granular and layered systems.

The first difference is related to the morphology of granular system.
Due to spherical grain shape the effective area of interaction
is small and it linearly depends on the grain size, $a$.
Therefore the intergrain exchange interaction in granular systems
grows linearly with $a$ in contrast to the MTJ, where interaction grows as $a^2$.

The second difference is the essential influence
of the Coulomb blockade effect on the hopping based exchange coupling.
In MTJ the Coulomb blockade is absent while in GFM the Coulomb interaction
suppresses the FM contribution to the hopping based magnetic intergrain coupling.

The third difference appears due to finite grain sizes.
The Coulomb based exchange interaction has an additional
contribution, $L^\mathrm{ex}_\mathrm{E_\mathrm c}$, appearing due to the
second term in Eq.~(\ref{Eq:CoulLeads}). This contribution does not
depend on the grain size $a$. On one hand the interaction area grows
linearly with $a$, and on the other hand this term is proportional to the charging energy $E_\mathrm c\sim 1/a$.

Thus, the total exchange interaction between magnetic grains
can not be extracted from the known result of interlayer exchange
coupling in MTJ by simple multiplication of the later by the grain or effective contact area.
\begin{figure}
\includegraphics[width=1\columnwidth]{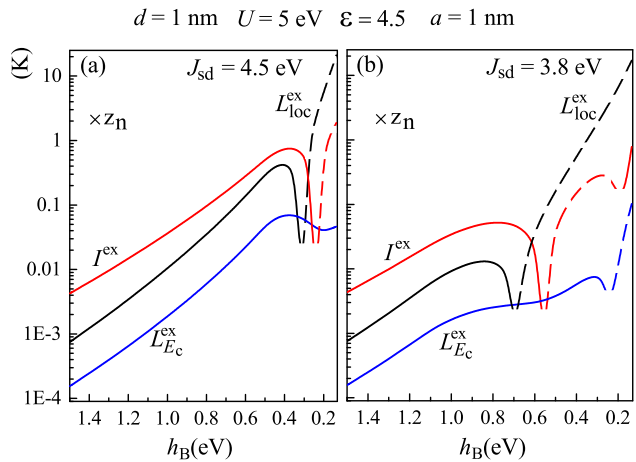}
\caption{(Color online) The intergrain exchange interaction (multiplied by the coordination number) as a function
of insulating barrier height $h_\mathrm B$ for $U=5$ eV, $\varepsilon=4.5$, $d=1$ nm, $a=1$ nm and (a) $\Ji=4.5$ eV, (b) $\Ji=3.8$ eV.
Black lines show $|L^\mathrm{ex}_\mathrm{loc}|$ (Eq.~(\ref{Eq:ExchContrCoulDelta})),
blue lines are for $|L^\mathrm{ex}_{E_\mathrm c}|$ (Eq.~(\ref{Eq:ExchContrCoulEc}))
and red lines are for $|I^\mathrm{ex}|$ (Eq.~(\ref{Eq:ExIns1})).
The y-axis has logarithmic scale. Dashed parts show the region where functions
$L^\mathrm{ex}_\mathrm{loc}$, $L^\mathrm{ex}_{E_\mathrm c}$ and $I^\mathrm{ex}$ are negative.} \label{Fig:JCvsEf1}
\end{figure}

\subsection{Comparison of different contributions to the Coulomb based exchange coupling in granular systems}

The Coulomb based intergrain exchange interaction has several contributions.
The first contribution, $I^\mathrm{ex}$, is due to the region between the grains.
In this region the Coulomb interaction can be considered as a long-range interaction.
The electric field of a point charge penetrates over the whole volume of the
insulator between the grains. This field is reduced by the dielectric between
the grains. Thus, the electron-electron interaction between the grains
depends on the dielectric constant of the insulating matrix, $\varepsilon$.
The second contribution appears due to the Coulomb interaction between
electrons inside the grains, $L^\mathrm{ex}$. It consists of two terms:
1) the short-range term in Eq.~(\ref{Eq:CoulLeads}), $L_\mathrm{loc}^\mathrm{ex}$, and
2) the size effect term, $L^\mathrm{ex}_{E_\mathrm c}$.
Terms $I^\mathrm{ex}$ and $L_\mathrm{loc}^\mathrm{ex}$ linearly grow with the grain size $a$.
The contribution $L^\mathrm{ex}_{E_\mathrm c}$ does not depend on
the grain size. Therefore the influence of this term increases with decreasing
the grain size $a$. However, our calculations show that even for
very small grains with $a\approx 1$ nm the contribution
$L^\mathrm{ex}_\mathrm{E_\mathrm c}$ is much smaller than two other contributions.
Figure~\ref{Fig:JCvsEf1} shows the behavior of these contributions to the Coulomb based
exchange interaction as a function of barrier height, $h_\mathrm B=\sqrt{-2m_\mathrm e E_\mathrm F/\hbar^2}$
(which is the difference between the energies of the insulator barrier and the Fermi level).
The curves are shown for very small grains, with grains diameter $2a=2$ nm.
Even in this case the contribution
$L^\mathrm{ex}_\mathrm{E_\mathrm c}$ exceeds two other contributions
only when $L_\mathrm{loc}^\mathrm{ex}$ or $I^\mathrm{ex}$ change its sign.
However, in this region the intergrain coupling due to the Coulomb
interaction is very small $\sim 10^{-2}$K. Thus, with a good accuracy we can
neglect the contribution $L^\mathrm{ex}_\mathrm{E_\mathrm c}$ in most cases.

Contributions $L_\mathrm{loc}^\mathrm{ex}$ and $I^\mathrm{ex}$ are comparable.
Figure~\ref{Fig:JCvsEf1} shows how these two contributions change their sign with
changing the barrier height, $h_\mathrm B$. For large barrier the interaction
is weak and positive (FM type), while for small barrier the interaction is negative
(AFM type). One can see that for large barrier the contribution due to the
intergrain region, $I^\mathrm{ex}$, exceeds contribution from the grains, $L_\mathrm{loc}^\mathrm{ex}$.
For small barrier the situation is the opposite, $L_\mathrm{loc}^\mathrm{ex}>I^\mathrm{ex}$.

Note that the contribution due to intergrain region
depends on the dielectric constant of the insulator,
$I^\mathrm{ex}\sim \varepsilon^{-1}$, while $L_\mathrm{loc}^\mathrm{ex}$ does not depend
on $\varepsilon$. Thus, changing the matrix dielectric constant, $\varepsilon$ one can
change the ratio of $L_\mathrm{loc}^\mathrm{ex}$ and $I^\mathrm{ex}$.
Figure~\ref{Fig:JCvsEf1} shows the case for $\varepsilon=4.5$, corresponding to Si insulator.
\begin{figure}
\includegraphics[width=1\columnwidth]{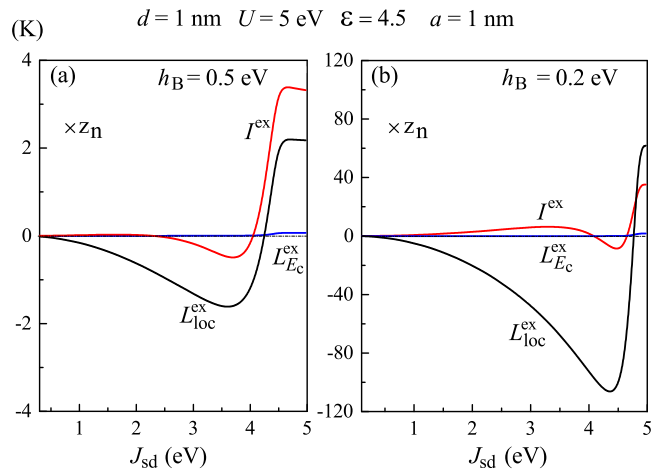}
\caption{(Color online) The intergrain exchange interaction as a function of
spin subband splitting, $\Ji$, for $U=5$ eV, $\varepsilon=4.5$, $d=1$ nm, $a=5$ nm,
and (a) $h_\mathrm B=0.5$ eV, (b) $h_\mathrm B=0.2$ eV.
Black lines show $L^\mathrm{ex}_\mathrm{loc}$ (Eq.~(\ref{Eq:ExchContrCoulDelta})),
blue lines are for $L^\mathrm{ex}_{E_\mathrm c}$ (Eq.~(\ref{Eq:ExchContrCoulEc}))
and red lines are for $I^\mathrm{ex}$ (Eq.~(\ref{Eq:ExIns1})).} \label{Fig:JCvsJ1}
\end{figure}

Figure~\ref{Fig:JCvsJ1} shows the dependence of three contributions
to the Coulomb based exchange interaction $L^\mathrm{ex}_\mathrm{loc}$, $L^\mathrm{ex}_{E_\mathrm c}$,
and $I^\mathrm{ex}$ on the spin subband splitting of electrons inside the grains, $\Ji$, for
$a=5$ nm grains. In this case the contribution $L^\mathrm{ex}_{E_\mathrm c}$ is negligible
in the whole range of parameters. The contribution due to
grains $L^\mathrm{ex}_\mathrm{loc}$ is negative (AFM) for
small splitting and positive (FM) for large splitting
(when only one spin subband is filled). The contribution coming from
the insulating region, $I^\mathrm{ex}$ changes its sign twice.
For small $\Ji$ the coupling is positive (FM), for intermediate $\Ji$ the contribution
is negative (AFM) and for large splitting $I^\mathrm{ex}>0$ (FM).

For large spin subband splitting (when only one subband is filled) and
for large barrier $h_\mathrm B$ the contribution $I^\mathrm{ex}$
exceeds the contribution coming from the grains
(Fig.~\ref{Fig:JCvsJ1}(a)). For small barrier the situation is the opposite.
For small splitting and for the case when both spin subbands
are filled ($\Ji<\Ef+U$) the contribution due to grains
exceeds the contribution due to the insulating
region ($|I^\mathrm{ex}|<|L^\mathrm{ex}_\mathrm{loc}|$).
In this region $L^\mathrm{ex}_\mathrm{loc}$ is of AFM type
and thus the whole Coulomb based coupling is of AFM type.

Note that for small barrier height
the Coulomb based coupling $|L^\mathrm{ex}_\mathrm{loc}|$ can be rather
large reaching 100 K. Thus, the intergrain Coulomb based exchange coupling
can be observed in experiment.

\subsection{Coulomb vs hopping based exchange interactions}

\begin{figure}
\includegraphics[width=1\columnwidth]{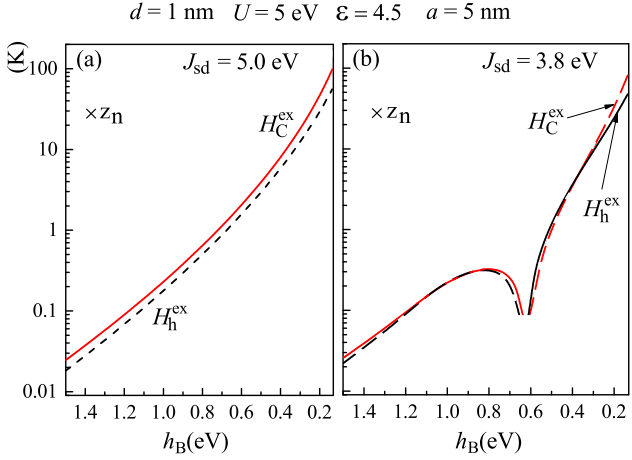}
\caption{(Color online) The intergrain exchange interaction (multiplied by the coordination number) as a function
of insulating barrier height $h_\mathrm B$ for $U=5$ eV, $\varepsilon=4.5$, $d=1$ nm, $a=5$ nm and (a) $\Ji=5$ eV, (b) $\Ji=3.8$ eV.
Black lines show the hopping based coupling $|H^\mathrm{ex}_\mathrm{h}|$ (Eq.~(\ref{Eq:ExFin})) and
red lines are for the Coulomb based coupling $|H^\mathrm{ex}_{\mathrm C}|$ (Eq.~(\ref{Eq:Ex2})). The y-axis has logarithmic scale. Dashed parts show the region where functions
$H^\mathrm{ex}_\mathrm{h}$ and $H^\mathrm{ex}_{\mathrm C}$ are negative.} \label{Fig:JHCvsEf1}
\end{figure}

Figure~\ref{Fig:JHCvsEf1} compares the hopping $H^\mathrm{ex}_\mathrm{h}$ and
the Coulomb $H^\mathrm{ex}_{\mathrm C}$ based exchange interactions
as a function of the barrier height $h_\mathrm B$ for the following parameters:
$U=5$ eV, $d=1$ nm, $a=5$ nm, $\varepsilon=4.5$ and (a) $\Ji=5.0$ eV, (b) $\Ji=3.8$ eV.
One can see that the Coulomb and the hopping based exchange couplings are comparable.
For large spin subband splitting, Fig.~\ref{Fig:JHCvsEf1}(a), the Coulomb based coupling exceeds
the hopping based coupling. For weak splitting ($\Ji<\Ef+U$) both contributions
change their sign. This happens almost for the same barrier height.
Contributions $H^\mathrm{ex}_\mathrm{h}$ and $H^\mathrm{ex}_{\mathrm C}$
have the opposite sign for almost all parameters.
For large spin subband splitting $H^\mathrm{ex}_\mathrm{h}$ is negative,
$H^\mathrm{ex}_\mathrm{h}<0$ (AFM) for any $h_\mathrm B$ while the Coulomb
based coupling is positive (FM). For
small splitting ($\Ji<\Ef+U$) the Coulomb based interaction $H^\mathrm{ex}_{\mathrm C}$
is positive for large barrier, and negative for small
barrier, while $H^\mathrm{ex}_\mathrm{h}$ shows the opposite behavior.

\begin{figure}
\includegraphics[width=1\columnwidth]{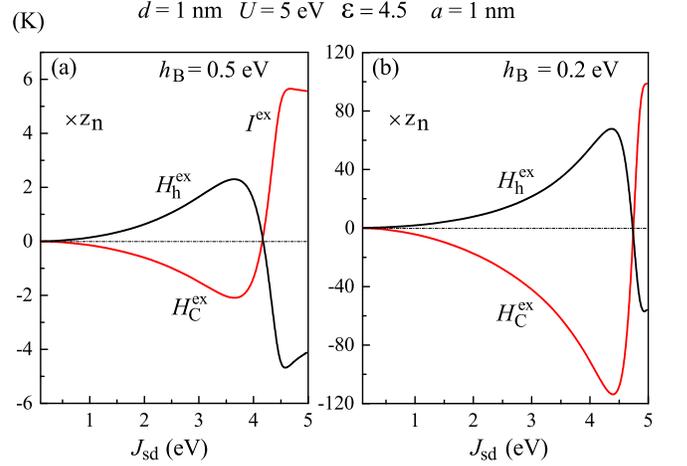}
\caption{(Color online) The intergrain exchange interaction as a function of
spin subband splitting, $\Ji$, for $U=5$ eV, $\varepsilon=4.5$, $d=1$ nm, $a=5$ nm,
and (a) $h_\mathrm B=0.5$ eV, (b) $h_\mathrm B=0.2$ eV.
Black lines show the hopping based coupling $H^\mathrm{ex}_\mathrm{h}$ (Eq.~(\ref{Eq:ExFin})) and
red lines are for the Coulomb based coupling $H^\mathrm{ex}_{\mathrm C}$ (Eq.~(\ref{Eq:Ex2})).} \label{Fig:JHCvsJ1}
\end{figure}

Figure~\ref{Fig:JHCvsJ1} shows the hopping based $H^\mathrm{ex}_\mathrm{h}$ and the Coulomb based $H^\mathrm{ex}_{\mathrm C}$ contributions to the total intergrain exchange interaction as a function of internal
spin subband splitting $\Ji$ for the following parameters: $U=5$ eV, $d=1$ nm,
$a=5$ nm, $\varepsilon=4.5$ and (a) $h_\mathrm B=0.5$ eV, (b) $h_\mathrm B=0.2$ eV.
One can see that both contributions are comparable and have the
opposite sign. For small splitting the hopping based contribution is positive (FM),
while the Coulomb based contribution is negative,
$H^\mathrm{ex}_{\mathrm C}<0$. For large splitting the situation is the opposite.

\subsubsection{A toy model}

The main feature of the hoping based and the Coulomb based contributions
is the sign change as a function of the barrier height $h_\mathrm B$ and the
spin subband splitting $J_\mathrm{sd}$. Moreover, one can see that the dependencies $H^\mathrm{ex}_\mathrm{h}$ and $H^\mathrm{ex}_\mathrm{C}$ on $h_\mathrm B$ and $J_\mathrm{sd}$ are quite similar but have the
opposite sign. The reason for such a similarity is related to the fact that
both $H^\mathrm{ex}_\mathrm{h}$ and $H^\mathrm{ex}_\mathrm{C}$ are defined by the density of
states in the vicinity of the Fermi surface. Consider the first term in
Eqs.~(\ref{Eq:ExHop0}) and (\ref{Eq:ExchContrCoulDelta}). The first integral
describes one of the hopping based contributions.
The second integral describes one of the Coulomb based contributions.
These two integrals are the most simple to analyse.
Due to the presence of the exponential factor, $e^{-2\varkappa d}$
only electrons in the vicinity of the Fermi surface contribute to
the integrals. We assume that the matrix elements do not depend on the electron energy
(besides the exponential factor). In this case we can estimate
\begin{equation}\label{Eq:TMhop1}
\begin{split}
H^{\mathrm{ex}}_{\mathrm{h0}}\!\sim\!\!V^-_{\mathrm F}\!\!N_--V^+_{\mathrm F}N_+-...
\end{split}
\end{equation}
and
\begin{equation}\label{Eq:TMExchContrCoulDelta}
\begin{split}
&L^\mathrm{ex}_{\mathrm{loc}}\sim\!\!\sum_{s,s'}\!\gamma(s,s')\!\!\int_0^{k_\mathrm F^s}\!\!\int_0^{k_\mathrm{F}^{s'}}\!\!\!\!\!dk_1 dk_2((k_\mathrm F^{s'})^2-k_2^2)((k_\mathrm F^s)^2-k_1^2)\times\\&\!\times\!(L_s e^{-2d\varkappa^s_{1}}\!+L_{s'}e^{-2d\varkappa^{s'}_{2}})=(N^0_-\! - \!N^0_+)(L_+ N_+\!-\!L_-N_-),\\
&N_s^0=\int_0^{k_\mathrm{F}^{s}}dk((k_\mathrm F^{s})^2-k^2),\\
&N_s=\int_0^{k_\mathrm{F}^{s}}dk((k_\mathrm F^{s})^2-k^2)e^{-2d\varkappa^s},\\
\end{split}
\end{equation}
where $V^{\pm}_F$ and $L_s$ are the parameters independent of integration variables. The key element of both the formulas is the integral of the form $\int ((k_\mathrm F)^2-k^2)e^{-2\varkappa d}dk$. This integral defines the number of electrons
participating in the exchange interaction. Equation~(\ref{Eq:TMhop1}) has only single integrals because this term is the first order perturbation theory correction to the system energy and it is proportional to the number of electrons in the system. Equation~(\ref{Eq:TMExchContrCoulDelta}) has double integrals since
it describes the many body interaction and it is proportional to the number
of electrons squared. The different spin subbands give contributions
to the exchange interaction of opposite sign.

For semi-metals (only one spin subband is
filled, $J_\mathrm{sd}>(U+E_\mathrm F)$) only the integrals over majority spin subband are survived.
Therefore, the majority spin subband defines the sign of the exchange interaction.
For small spin subband splitting, $J_\mathrm{sd}\ll E_\mathrm F$ (and $\varkappa_0\ll k_\mathrm F$)
we have
\begin{equation}\label{Eq:Auxiliary1}
\int ((k^s_\mathrm F)^2-k^2)e^{-2\varkappa^s d}dk\sim\frac{\varkappa_0^3}{dk^s_\mathrm F}e^{-2\varkappa_0 d}.
\end{equation}
This result means that the spin subband with
higher density of states at the Fermi surface (higher $k_\mathrm F$)
gives the smaller contribution to the exchange interaction meaning
that at small $J_\mathrm{sd}$ the minority spin subband defines
the sign of the exchange interaction. This causes the sign change of the
exchange coupling at a certain $J_\mathrm{sd}$.
To estimate the transition point we estimate the
integral $\int ((k^s_\mathrm F)^2-k^2)e^{-2\varkappa^s d}dk$ at small Fermi
momentum $k_F^+\ll \varkappa_0$. The estimate in Eq.~(\ref{Eq:Auxiliary1}) does
not work in this limit ($k_\mathrm F\to 0$). We have $\int ((k^+_\mathrm F)^2-k^2)e^{-2\varkappa d}dk\sim (k^+_\mathrm F)^3e^{-2\varkappa_0d}$ and $\int ((k^-_\mathrm F)^2-k^2)e^{-2\varkappa d}dk\sim(\varkappa_0^3)/(dk^-_\mathrm F)e^{-2\varkappa_0 d}$. The exchange interaction changes its sing when the integrals for both spin subbands
are equal. This point is defined by the condition $\varkappa_0^3\approx d k^-_\mathrm F (k^+_\mathrm F)^3$.
Usually, $\varkappa_0^2\ll E_\mathrm F$ and therefore, the transition appears
close to the point $k^+_\mathrm F=0$, i.e. close to the case of
semimetal ($J_\mathrm{sd}\approx(U+E_\mathrm F)$). This is in agreement
with our calculations. The condition also shows that the sign change
appears with varying the barrier heigh $h_\mathrm B$, which
is also in agreement with our calculations. This toy model explains the
behavior of the exchange interaction and the reason for
similarity between the Coulomb and the hopping based exchange contributions.

\subsection{Total exchange interaction}

In granular systems the Coulomb and the
hopping based exchange interactions compete with each other. These two contributions
have the opposite sign for almost all parameters.
\begin{figure}
\includegraphics[width=1\columnwidth]{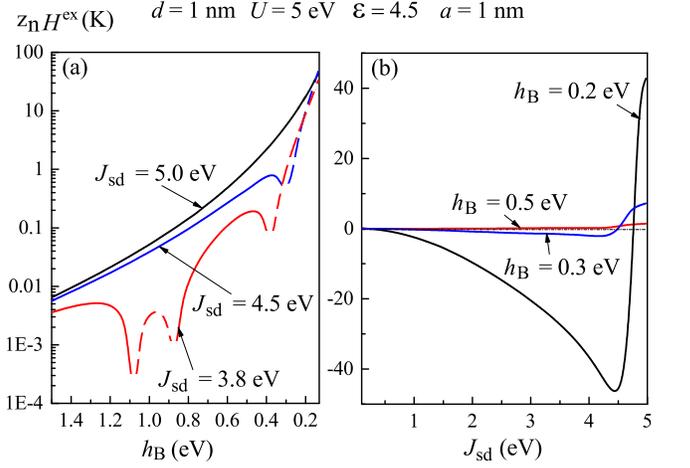}
\caption{(Color online) Total intergrain exchange interaction $H^\mathrm{ex}$ (Eq.~(\ref{Eq:ExTot})) as a function of
(a) the barrier height $h_\mathrm B$, and (b) spin subband splitting, $\Ji$, for $U=5$ eV, $\varepsilon=4.5$, $d=1$ nm, $a=5$ nm. In plot (a) the y-axis has logarithmic scale. Dashed parts show the region where function
$H^\mathrm{ex}$ is negative.} \label{Fig:JTotvsJandEf1}
\end{figure}

Figure~\ref{Fig:JTotvsJandEf1} shows the total intergrain exchange
interaction, $H^\mathrm{ex}$ as a function of (a) the barrier height $h_\mathrm B$,
and (b) the spin subband splitting, $\Ji$, for $U=5$ eV, $\varepsilon=4.5$, $d=1$ nm, $a=5$ nm.
The sign and the magnitude of the total exchange interaction depends on the value of
spin subband splitting, $\Ji$ and the barrier height, $h_\mathrm B$.
For small splitting $\Ji$ the coupling is AFM while for large
splitting it is FM. Depending on $J_\mathrm{sd}$ the coupling changes
its sign one or three times. Due to the competition between the Coulomb and
the hopping mechanisms the magnitude of the total exchange interaction
is smaller than the magnitude of the Coulomb based contribution.

Note that both the Coulomb and the hopping based
contributions depend on the dielectric permittivity of the
insulating matrix. The Coulomb contribution can be written as
\begin{equation}\label{Eq:ExCoulEps}
H^\mathrm{ex}_\mathrm C=L^\mathrm{ex}_{\mathrm{loc}}+\frac{I^\mathrm{ex}_1}{\varepsilon},
\end{equation}
where $I^\mathrm{ex}_1$ is the Coulomb based exchange coupling
inside the insulator with $\varepsilon=1$.
Note that $I^\mathrm{ex}_1$ can be either positive or negative depending on the system parameters.
The dependence of the hopping
contribution $H^\mathrm{ex}_\mathrm{h}$ on the dielectric constant
is more complicated (see Ref.~[\onlinecite{Bel2016ExGr}]). Approximately it can be written as
\begin{equation}\label{Eq:ExHopEps}
H^\mathrm{ex}_\mathrm h=H^\mathrm{ex}_{\mathrm{h}0}+H^\mathrm{ex}_{\mathrm h 1}\left(\!\!1-\sqrt{\frac{ d \sqrt{2m}\tilde\epsilon_{\mathrm c}}{\gamma\hbar\sqrt{h_\mathrm B}}}\mathrm{arctan}\left(\sqrt{\frac{\gamma\hbar\sqrt{h_\mathrm B}}{ d \sqrt{2m}\tilde\epsilon_{\mathrm c}}}\right)\!\!\right),
\end{equation}
where $\gamma\approx3.43$ and $H^\mathrm{ex}_{\mathrm h 1}>0$. The
dielectric permittivity in this equation enters through the effective
charging energy, $\tilde\epsilon_\mathrm c\sim 1/\varepsilon$,
for simplicity we omit the difference between $\varepsilon_\mathrm{eff}$
and $\varepsilon$. The second term in Eq.~(\ref{Eq:ExHopEps}) increases
with increasing $\varepsilon$. This is in contrast to the Coulomb based
coupling. Also, we note that $\tilde\epsilon_\mathrm c$ depends on the grain size, $a$.
Decreasing the grain size leads to the enforcement of the Coulomb blockade
effect making $H^\mathrm{ex}_\mathrm h$ more sensitive to variation of $\varepsilon$.
\begin{figure}
\includegraphics[width=1\columnwidth]{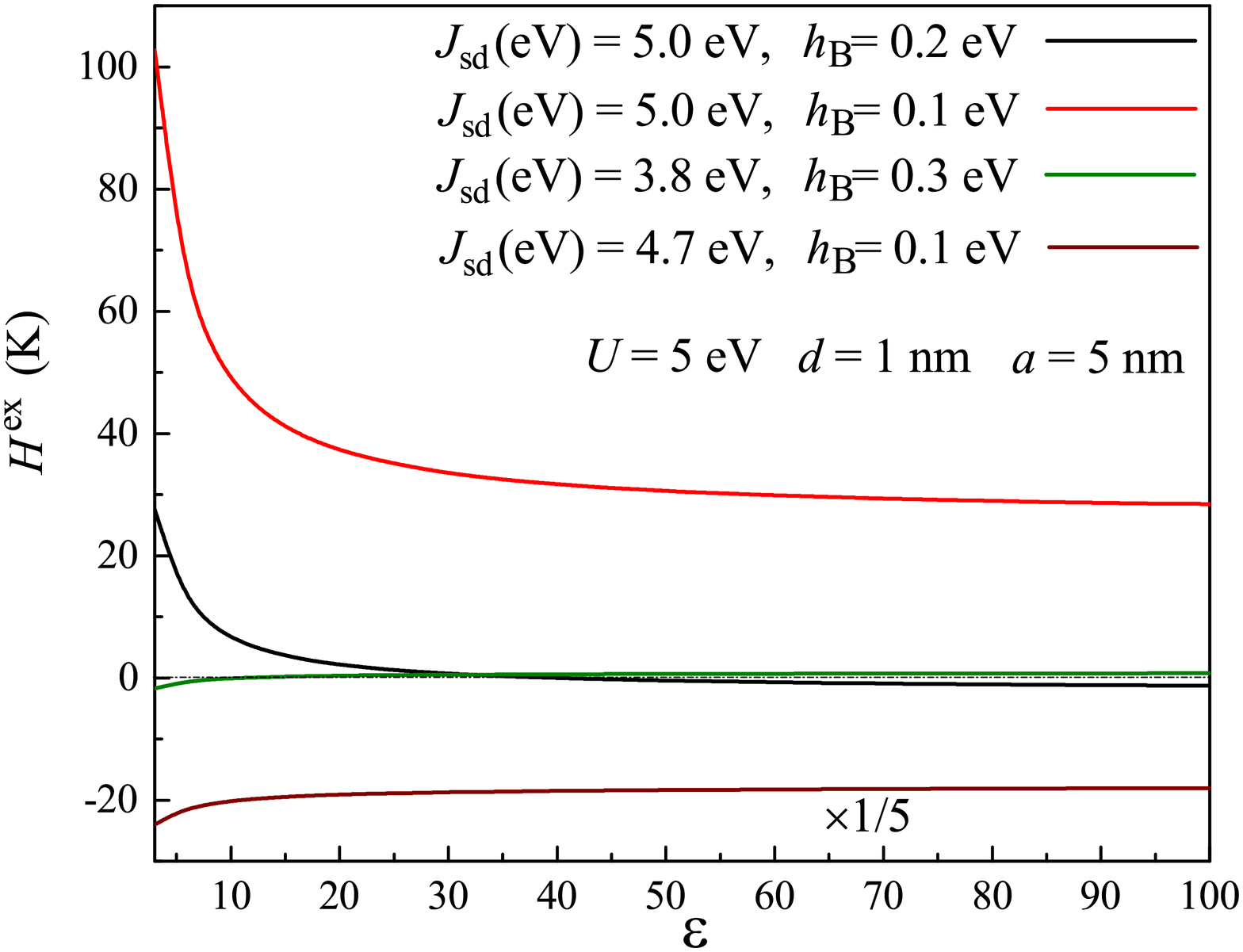}
\caption{(Color online) Total intergrain exchange coupling, $H^\mathrm{ex}$ in Eq.~(\ref{Eq:ExTot}) as a function of dielectric
permittivity of the insulating layer, $\varepsilon$, for $U=5$ eV, $d=1$ nm, $a=5$ nm and
different spin subband spitting, $\Ji$, and barrier height $h_\mathrm B$. The brown curve divided by 5.} \label{Fig:JTotvsEps1}
\end{figure}

Using Eqs.~(\ref{Eq:ExCoulEps}) and (\ref{Eq:ExHopEps}) we can write
\begin{equation}\label{Eq:ExTotEps}
\begin{split}
&H^\mathrm{ex}=H^\mathrm{ex}_{0}+\\
&+\frac{I^\mathrm{ex}_1}{\varepsilon}+H^\mathrm{ex}_{\mathrm h 1}\left(\!\!1-\sqrt{\frac{ d \sqrt{2m}\tilde\epsilon_{\mathrm c}}{\gamma\hbar\sqrt{h_\mathrm B}}}\mathrm{arctan}\left(\sqrt{\frac{\gamma\hbar\sqrt{h_\mathrm B}}{ d \sqrt{2m}\tilde\epsilon_{\mathrm c}}}\right)\!\!\right).
\end{split}
\end{equation}
The second and the third terms have opposite dependence on $\varepsilon$. Figure~\ref{Fig:JTotvsEps1} shows the dependence of the total exchange interaction $H^\mathrm{ex}$ on the dielectric permittivity of the insulating
matrix for various parameters. In most cases the Coulomb based contribution $H^\mathrm{ex}_\mathrm C$,
Eq.~(\ref{Eq:ExTotEps}), is
the largest. For positive $I^\mathrm{ex}_1$ the total exchange coupling
decreases with increasing $\varepsilon$. One can see that $H^\mathrm{ex}_{0}$
can be either positive (red curve) or negative (black curve). For positive $H^\mathrm{ex}_{0}$
the exchange coupling is always positive (FM) and decreases
with increasing the matrix dielectric constant. For negative $H^\mathrm{ex}_{0}$
the coupling changes its sign with increasing $\varepsilon$.
For small dielectric constant, $H^\mathrm{ex}_\mathrm C$ is of FM type, and it
becomes AFM for large dielectric constants. The total coupling decreases
three times (from 100 K to 30 K) with increasing the dielectric constant.

For some parameters the hopping based coupling is the dominant
contribution to $\varepsilon$-dependence of the total exchange interaction, $H^\mathrm{ex}$ (green line).
In this case the coupling grows with $\varepsilon$. For small
dielectric constant the coupling is of AFM type. It becomes positive (FM)
with increasing the dielectric constant.

For negative $I^\mathrm{ex}_1$ the total coupling is negative
and increases (the absolute value decreases)
with increasing the dielectric constant (brown curve in Fig.~\ref{Fig:JTotvsEps1}).
In this case both contributions contribute in the same direction.
Thus, changing system parameters one can have different dependencies
of the exchange coupling on $\varepsilon$ in granular systems.

The fact that the total intergrain exchange interaction depends
on the dielectric constant can be used to realize the magneto-electric
coupling in granular systems. This effect was semi-phenomenologically predicted in Refs.~[\onlinecite{Bel2014ME,Bel2014ME1,Bel2014ME2}], where it was shown that
if one can control the dielectric properties of the matrix with external
electric field than one can control the intergrain exchange coupling
and therefore the magnetic state of the granular magnet.
For example, the FE matrix can be used for this purpose.
It is known that the dielectric permittivity of FEs depends
on the electric field. Applying the electric field to the system
with magnetic grains being placed into FE matrix one can
change its magnetic state.

\section{Validity of our model}\label{Sec:Val}

Below we discuss several assumptions and approximations of our theory.

1) Above we introduce two dielectric constants:
the real constant $\varepsilon$ and the effective constant $\varepsilon_\mathrm{eff}$.
The constant $\varepsilon$ defines the screening of electric field
in the region between the grains (insulating matrix). This constant
governs the exchange coupling based on the Coulomb interaction.
The electric field involved in this interaction exists
only in the small region between the grains. The effective
dielectric constant $\varepsilon_\mathrm{eff}$ describes the
long-range screening on the scale of many grains. A charged
grain creates a field penetrating into volume of many grains.
Therefore the effective dielectric constant $\varepsilon_\mathrm{eff}$
includes the screening properties of both the matrix and the grains.
Thus, the charging energy and the hopping based exchange coupling
depend on the dielectric properties averaged over large volume,
while the Coulomb based coupling depends on the dielectric
properties of a small intergrain region. A qualitative difference
between these two constants may appear in the system
with magnetic grains being placed on a substrate with variable
dielectric constant. Such a substrate will influence the charging
energy (see Ref.~[\onlinecite{Bel2014GFE1,Bel2014GFE2}])
and therefore the hopping based exchange coupling. However,
it will not influence the Coulomb based exchange coupling.

2) We propose to use FE as an insulating matrix with variable
dielectric constant. To observe the intergrain exchange coupling
in experiment the intergrain distance should be of the
order of 1 nm. The properties of such thin FE films are not
well known at this time. However, it is known that
FE properties degrade with decreasing of FE thickness~[\onlinecite{Frid2006rev,Frid2010rev}].
For each particular FE there is a critical thickness at which FE properties disappear.
At the same time the mono-atomic layer FEs also exist~[\onlinecite{Frid2006rev,Frid2010rev}].
The FE properties of a dense granular material with magnetic inclusions
are not studied at all. This question requires further investigation.

3) Following Ref.~[\onlinecite{Bel2016ExGr}] we do not take into account the inelastic scattering and tunneling.

4) When calculating the Coulomb based contribution to the total
exchange coupling we use the approach of Ref.~[\onlinecite{Beloborodov2016ExLayer}] which
was developed for infinite layered system. The grains form a capacitor
with finite lateral size with electric charge being  localized in between the
capacitor surfaces (grains) and inside the grains. The
charge is localized in the area $S_\mathrm c$. In our calculations we assume
that electric field is localized between the leads only. Such an approximation
is valid when the lateral size of the capacitor is much larger than
capacitor thickness. We calculated numerically the energy of a finite flat
capacitor with uniformly distributed positive charge inside the capacitor
and negatively charged surfaces, such that the whole system is neutral.
The capacitor area is $S_\mathrm c$. The energy of the capacitor is $W^{\mathrm{fc}}$.
We compare the energy with the energy of the area $S_\mathrm c$ of an
infinite flat capacitor $W^{\mathrm{ic}}$,
$W^{\mathrm{ic}}<W^{\mathrm{fc}}$. The difference between $W^{\mathrm{ic}}$
and $W^{\mathrm{fc}}$ is of order of $d/\sqrt{S_\mathrm c}$, where $d$ is
the capacitor thickness. Thus, the matrix element of the exchange interaction
is overestimated. The error grows with decreasing the grain size.

5) We also assume that the leads are perfect metals
meaning that they totally screen the electric field. In fact,
the electric potential created by a point charge located in a
metal decays exponentially with distance, $\sim e^{-r/\lambda_\mathrm{TF}}/r$,
where $\lambda_\mathrm{TF}$ is the Thomas-Fermi length. The field of a point
change outside the metal surface also penetrates into the metal by
the distance of the order of Thomas-Fermi length. The Thomas-Fermi
length is of the order of 0.05 nm and is much smaller than
the characteristic length scales of the decay of electron
wave function $\varkappa_0$ and the insulator thickness $d$.
Our approach is valid for $\lambda_\mathrm{TF} < \mathrm{min}(\varkappa_0, d)$.

\section{Conclusion}

We developed the theory of the intergrain exchange interaction in the
system of two metallic magnetic grains embedded into an insulating matrix by
taking into account the magnetic coupling due to Coulomb interaction
between electrons. The basic idea is the following:
electrons wave functions located at different grains are overlapped.
In combination with weak screening of electric field inside the
insulator these electrons experience the indirect spin-dependent Coulomb interaction
leading to interlayer magnetic coupling. The Coulomb based exchange interaction
complements the exchange interaction due to virtual electron hopping between
the grains. We showed that the Coulomb and the hopping based exchange interactions are
comparable. For most of the parameters these two contributions have the
opposite sign and therefore compete with each other. 

We showed that many-body effects lead to new phenomena in
magnetic exchange coupling. In particular, the exchange coupling
depends not only on the barrier height
and thickness of the insulating matrix but also on the dielectric properties
of this matrix. In granular systems both the hopping and the Coulomb
based exchange coupling depend on the dielectric constant of the insulating
matrix. This dependence appears due to many-body effects.
We showed that hopping based exchange interaction depends on the matrix
dielectric constant due to the Coulomb blockade effect controlling
virtual electron hopping between the grains. The larger the dielectric
constant the smaller the Coulomb blockade thus the stronger the exchange
coupling. The Coulomb based exchange coupling depends on the dielectric
constant $\varepsilon$ - decreasing with increasing $\varepsilon$.
Both the hopping and the Coulomb based exchange interactions have
terms which do not depend on the matrix dielectric constant. These terms
can be either FM or AFM type. The combination of three different
contributions to the total exchange coupling results in a complicated dependence
of the total magnetic intergrain exchange on $\varepsilon$ and other parameters
of the system. Increasing $\varepsilon$ one can have the FM - AFM or
AFM-FM transitions. For certain parameters no transition is possible,
however the exchange coupling varies by three times with
increasing the dielectric constant.

We showed that the intergrain exchange interaction strongly depends on
system parameters such as Fermi level, internal spin subband splitting,
the height of the insulating barrier and the grain size. The dependence
on the grain size is almost linear due to spherical shape of the grains.
The contact area of two grains linearly depends on the grain size in contrast to
layered system, where the exchange coupling increases as the surface area.
Depending on the Fermi level and the spin subband splitting the
intergrain exchange coupling can be either positive (FM) or negative (AFM).
For small barrier height the coupling can be rather strong even for 5 nm grains
reaching 100 K if the spin subband splitting is large enough.

\section{Acknowledgements}

This research was supported by NSF under Cooperative Agreement Award EEC-1160504,
the U.S. Civilian Research and Development Foundation (CRDF Global) and NSF PREM Award. O.U. was supported by Russian Science Foundation (Grant  16-12-10340).

\bibliography{Exchange}

\end{document}